\documentclass[conference]{IEEEtran}
\IEEEoverridecommandlockouts
\usepackage{cite}
\usepackage{amsmath,amssymb,amsfonts}
\usepackage{algorithmic}
\usepackage{graphicx}
\usepackage{microtype}
\usepackage{textcomp}
\usepackage{xcolor}
\usepackage{listings}
\usepackage{setspace}
\usepackage{dblfloatfix}
\usepackage{float}
\usepackage[all]{nowidow}
\usepackage{url}
\def\BibTeX{{\rm B\kern-.05em{\sc i\kern-.025em b}\kern-.08em
    T\kern-.1667em\lower.7ex\hbox{E}\kern-.125emX}}

\makeatletter
\def\lst@makecaption{%
	\def\@captype{figure}%
	\@makecaption
}
\makeatother

\begin{document}

\title{{A QUIC(K) Way Through Your Firewall?}\\
\Large{Exploring Bypass Capabilities and Mitigation Strategies of Stateful Firewalls\\in the Presence of HTTP/3 / QUIC}
}

\author{
\IEEEauthorblockN{Konrad Yuri Gbur}
\IEEEauthorblockA{
\textit{Technische Universität Berlin}\\
konrad.y.gbur@campus.tu-berlin.de}
\and
\IEEEauthorblockN{Florian Tschorsch}
\IEEEauthorblockA{
	\textit{Technische Universität Berlin}\\
	florian.tschorsch@tu-berlin.de}
}

\maketitle
\thispagestyle{plain}
\pagestyle{plain}

\begin{abstract}
	The QUIC protocol is a new approach to combine encryption and transport layer stream abstraction into one protocol to lower latency and improve security. However, the decision to encrypt transport layer functionality may limit the capabilities of firewalls to protect networks. To identify these limitations we created a test environment and analyzed generated QUIC traffic from the viewpoint of a middlebox. This paper shows that QUIC indeed exposes traditional stateful firewalls to UDP hole punching bypass attacks. On the contrary we show the robustness against censorship of QUIC through the encrypted transport layer design and analyze the capabilities to re-gain stateful tracking capabilities by deep packet inspection of the few exposed QUIC header fields.
\end{abstract}

\section{Introduction}
Since the web was first proposed in 1989, the primary transport protocol used for all the Hypertext Transport Protocol~(HTTP) traffic has been the Transmission Control Protocol~(TCP). Providing a reliable data stream abstraction directly in the transport layer of the ISO/OSI network protocol stack, TCP was more suitable than the connection-less User Datagram Protocol~(UDP) for providing documents via HTTP~\cite{ietfHTTP1,berners-lee1997}.

With the web transforming from hosting static documents only to hosting entire security-critical web applications such as online banking, encrypted communication became a core requirement. As the original Internet protocol stack used for the web did not include any capabilities for encryption, the Secure Socket Layer~(SSL), now called Transport Layer Security~(TLS), was introduced. Implementing the wanted encryption capabilities, TLS requires additional handshakes, resulting in two more round-trip times~(RTT) to set up a secure connection~\cite{ietfHTTPTLS}.

\begin{figure}[tb]
	\centerline{\includegraphics[width=\columnwidth]{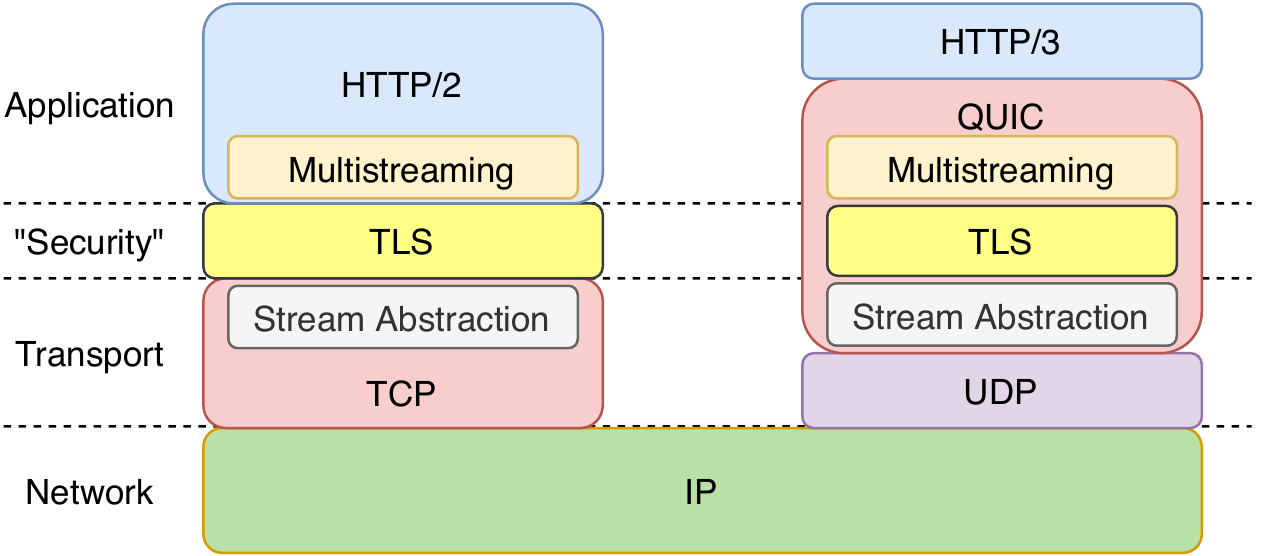}}
	\caption{Protocol stacks of HTTP/2 over TCP/TLS and HTTP/3 over QUIC based on~\cite{cui2017} and~\cite{langley2017}. The original ISO/OSI network stack did not include a specific security layer but TLS is usually regarded as a separate layer.}
	\label{fig:protocol_stacks}
	\vspace{-1ex}
\end{figure}

The extra amount of RTTs became an issue for increasingly popular real-time applications in high latency networks. The most prominent solution approach is the Quick UDP Internet Connection~(QUIC), laying the foundation for the new HTTP/3 standard~\cite{langley2017, lychev2015}. The differences between the HTTP/2 and HTTP/3 protocol are shown in Figure~\ref{fig:protocol_stacks}. It shows that QUIC combines the transport layer stream abstraction, the encryption handling and parts of the application layer into one layer. By combining these components, QUIC is able to eliminate the two additional RTTs introduced by TLS~\cite{langley2017}. QUIC can further be configured to utilize the new TLS 1.3 0-RTT handshake for known endpoints, completely eliminating the RTTs for the connection setup. At the time of writing, the Internet Engineering Task Force~(IETF) standardizes QUIC and HTTP/3 which could lead to a change of the web's core protocol from TCP to UDP in the near future~\cite{ietfQUIC29,ietfQUICTLS29,ietfHTTP329}.

However, besides the advantages through QUIC, the change from TCP to UDP may inherit some security challenges for firewalls. Stateful firewalls rely on connection tracking to distinguish which packets belong to a specific client-server communication. For example, a web server firewall will only allow incoming traffic on the HTTP(S) ports 80 and 443 to establish a connection, as there is no need for a web server to initiate connections to hosts outside the intranet in most cases. To distinguish connections, it is not sufficient to rely on IP addresses and ports only. The firewalls have to interpret more of the packets content to reconstruct the stream abstraction. In case of TCP, this can be implemented by analyzing the unencrypted \texttt{SYN}, \texttt{ACK}, \texttt{FIN} and \texttt{RST} flags in the TCP header which indicate handshakes, flow directions and teardowns of connections~\cite{goralski2009}.

Since QUIC merges the stream abstraction and the encryption layer with the goal to provide as much privacy as possible and reduce latency, the relevant header fields for stateful connection tracking are encrypted as well. Only a minimum of header fields are exposed for analysis through middleboxes~\cite{langley2017, ietfQUIC29}. Besides the limited insight a firewall could get from the few unencrypted header fields, many stateful firewall implementations, may not even analyze protocol information above the transport layer in the first place. QUIC is built on top of UDP to provide compatibility with old network stacks and it is not regarded as a transport layer protocol in a traditional view~\cite{langley2017}.

Firewalls built to perform a deep packet inspection above the transport layer have to parse the packets and decide heuristically which protocol is used in a general case because the transport layer does not contain any information about the higher layer protocols. This could impact the performance of the firewall and might not be an option in many network scenarios~\cite{kumar2006}. Decker~\cite{decker2020} has already shown that existing intrusion detection systems~(IDS) and intrusion prevention systems~(IPS) are not able to detect malicious QUIC traffic and many providers of firewalls recommend not to use QUIC until sophisticated methods for detection are developed.

In this paper, we analyze how traditional transport layer based stateful firewalls like the linux \emph{conntrack} module cope with QUIC. Our contributions are the following:
\begin{itemize}
	\item We show that firewalls handling QUIC are prone to UDP hole punching, a technique that is well known and has been documented for years but so far has a limited effect in the real world for web servers protected by firewalls.
	\item We demonstrate the threats of UDP hole punching by creating a reverse shell that would not be similarly possible in a TCP/TLS scenario.
	\item Furthermore, we show that one of the reasons for this attack being possible with QUIC is also the reason for its resistance to censorship.
	\item We analyze which information are exposed by QUIC and can be used for stateful connection tracking without reducing the censorship resistance.
\end{itemize}

\begin{figure}[tb]
	\centerline{\includegraphics[width=\columnwidth]{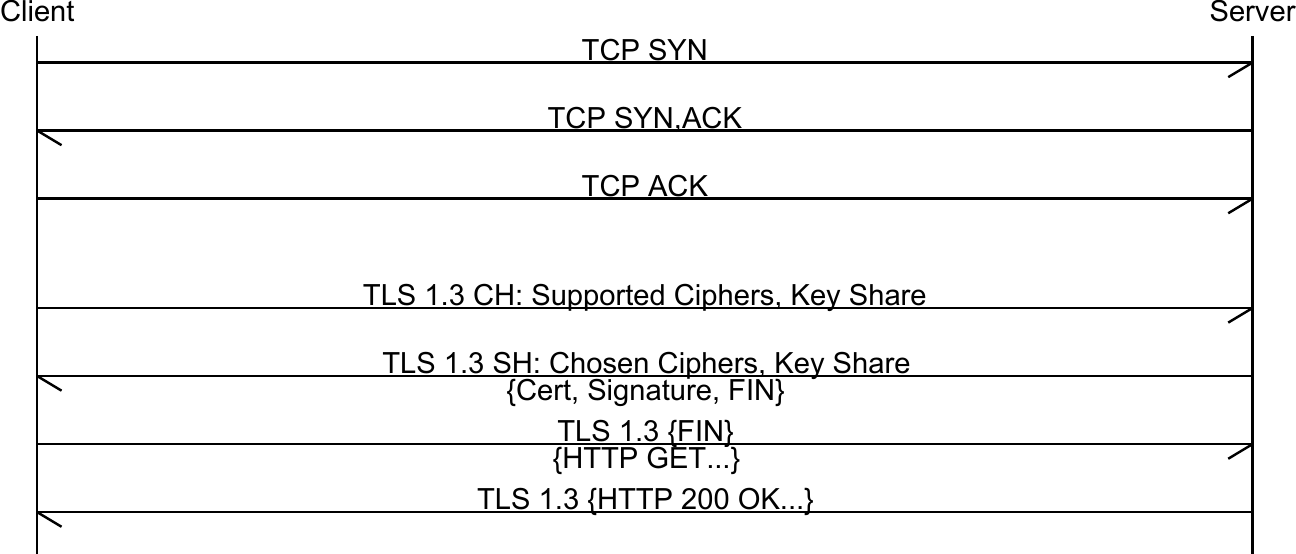}}
	\caption{TCP/TLS 1.3 Handshakes for encrypted communication.}
	\label{fig:tcp_tls_handshake}
\end{figure}

The paper is structured in the following way: Section~\ref{sec:background} provides a short introduction into the concepts of stream abstraction and encryption with TLS. Our testing setup is explained in Section~\ref{sec:setup}. In Section~\ref{sec:conntrack}, we analyze how statefulness is implemented for UDP packets in \emph{iptables} utilizing \emph{conntrack} and compare it to TCP in perspective of firewall bypass capabilities. First approaches to stateful tracking capabilities in QUIC in order to mitigate UDP hole punching are explored in Section~\ref{sec:quic_http3_dissection} by dissecting the network traffic and correlating it with the QUIC specification. This section also shows how the QUIC protocol design achieves a resistance against reset attacks as they exist in TCP. Section~\ref{sec:futurework} contains a discussion about the topic and implications for future work before we give a conclusion in Section~\ref{sec:conclusion}.

\section{Background}
\label{sec:background}
Since QUIC is directly built upon the encryption capabilities provided by TLS 1.3, this section introduces the basics of how the TLS handshake, shown in Figure~\ref{fig:tcp_tls_handshake}, works. The client starts initiating the TCP 3-way handshake with a server that is listening on a specific port. After the stream is set up, the client sends a \emph{Client Hello}~(CH) message to the server along with the supported cipher suites and the necessary key information for one cipher suite (\emph{key share}) that is most likely supported by the server. The choice of algorithm could be preset or based on a guess performed in the client implementation. On receiving the CH, the server answers with a \emph{Server Hello}~(SH) containing the chosen cipher suite and the \emph{key share} for the server-side in plain text. Furthermore, the server certificate, the signature and a hashed transcript of the setup so far, called the \emph{Finished Message} (\texttt{FIN}), are transmitted in the same message but already encrypted. If the data provided by the server is trusted by the client and the validity of the \texttt{FIN} message is confirmed, the client also sends an encrypted \texttt{FIN} message finalizing the cryptographic handshake. Together with the \texttt{FIN} message from the client, the first application data can already be transmitted~\cite{ietfTLS1.3, ietfQUICTLS29}.

\begin{figure}
	\begin{lstlisting}[language=bash]
# Generated by iptables-save v1.8.4
*filter
:INPUT DROP [581:48486]
:FORWARD DROP [0:0]
:OUTPUT DROP [27:3466]
-A INPUT -p tcp -m tcp --dport 443 -m conntrack --ctstate NEW,RELATED,ESTABLISHED -j ACCEPT
-A INPUT -p udp -m udp --dport 443 -m conntrack --ctstate NEW,RELATED,ESTABLISHED -j ACCEPT
-A OUTPUT -p tcp -m tcp --sport 443 -m conntrack --ctstate RELATED,ESTABLISHED -j ACCEPT
-A OUTPUT -p udp -m udp --sport 443 -m conntrack --ctstate RELATED,ESTABLISHED -j ACCEPT
COMMIT
	\end{lstlisting}
	\caption{The \emph{iptables} rules deployed on the server.}
	\label{fig:lst:iptables_rules}
\end{figure}

\section{Testing Setup}
\label{sec:setup}
To test how QUIC traffic is working, a simple point-to-point topology was chosen with a local firewall on the server. We have used the QUIC implementation \emph{quiche} by Cloudflare~\cite{github_quiche} which provides a patch for \emph{Nginx} to support HTTP/3 and can be integrated into the \emph{curl}\cite{github_curl} web client tool. The QUIC implementation by Cloudflare is based on the IETF QUIC and HTTP/3 drafts in revision 29~\cite{ietfQUIC29, ietfHTTP329}. Since \emph{curl} also supports the current HTTP standard over TCP/TLS it can be used for direct comparison of the TCP/TLS and QUIC protocol stacks. Both hosts are running as virtual machines~(VM) in the same local network. The client is an Ubuntu 20.04 64-bit with the local IP 192.168.79.132 and the server is an Ubuntu Server 20.04 reachable over the IP 192.168.79.128. In the following sections, the IP addresses are indicated by the last octet only. Detailed technical information about the setup of \emph{curl} on the client and the \emph{Nginx} server as well as the reset attack script used later can be found in our GitHub repository~\cite{github_setup}.

Figure~\ref{fig:lst:iptables_rules} lists the \emph{iptables} rules used on the server. The firewall allows incoming TCP and UDP traffic only for the destination port 443 and allows outgoing packets from this port only for established connections or related packets. The connection state is tracked with the \emph{conntrack} module which also provides an API to access the tables via the command line tool also called \emph{conntrack}. All information about the connection state was retrieved from this tool.

\section{Stateful Connection Tracking}
\label{sec:conntrack}
Stateless firewalls treat all incoming packets individually and compare them to a list of rules. Packets are interpreted as 5-tuples consisting of protocol, source IP address, destination IP address, source port, and destination port. There exists a set of rules for in- and outbound traffic, with each rule matching a number of entries in the 5-tuple for a packet. This concept is not sufficient for real-world applications in most cases. In our testing topology, a stateless firewall would not be able to allow outbound traffic only if inbound traffic initiated the connection beforehand (and vice-versa), because there is no possibility to ``remember'' the previous inbound connection. Furthermore, rules for static ports do not work, since source ports are usually chosen randomly by the client~\cite{goralski2009}.

To cope with random source ports, stateful firewalls maintain a state table. This table stores 5-tuples of processed packets together with the state a packet represents. With this table it is possible to allow packets only if there was a connection setup from the other direction beforehand. This concept works especially well for TCP, where all the stream abstraction is present unencrypted in the transport layer headers and can be used to track connections. Concerning UDP, firewalls are able to track packets in a table but they can only guess if returning packets with the same 5-tuple belong to the same connection. All table entries are paired with a time-to-live~(TTL) field that is reset or changed with each update. If an entry times out, the entry is destroyed~\cite{goralski2009}. To see how exactly stateful firewalls (i.e., \emph{conntrack}) exactly cope with TCP and UDP packets, the firewall tracking mechanisms for both protocols and the resulting implications for the security of the firewall are examined in more detail in the following paragraphs.

\begin{figure}[tb]
	\centerline{\includegraphics[width=\columnwidth]{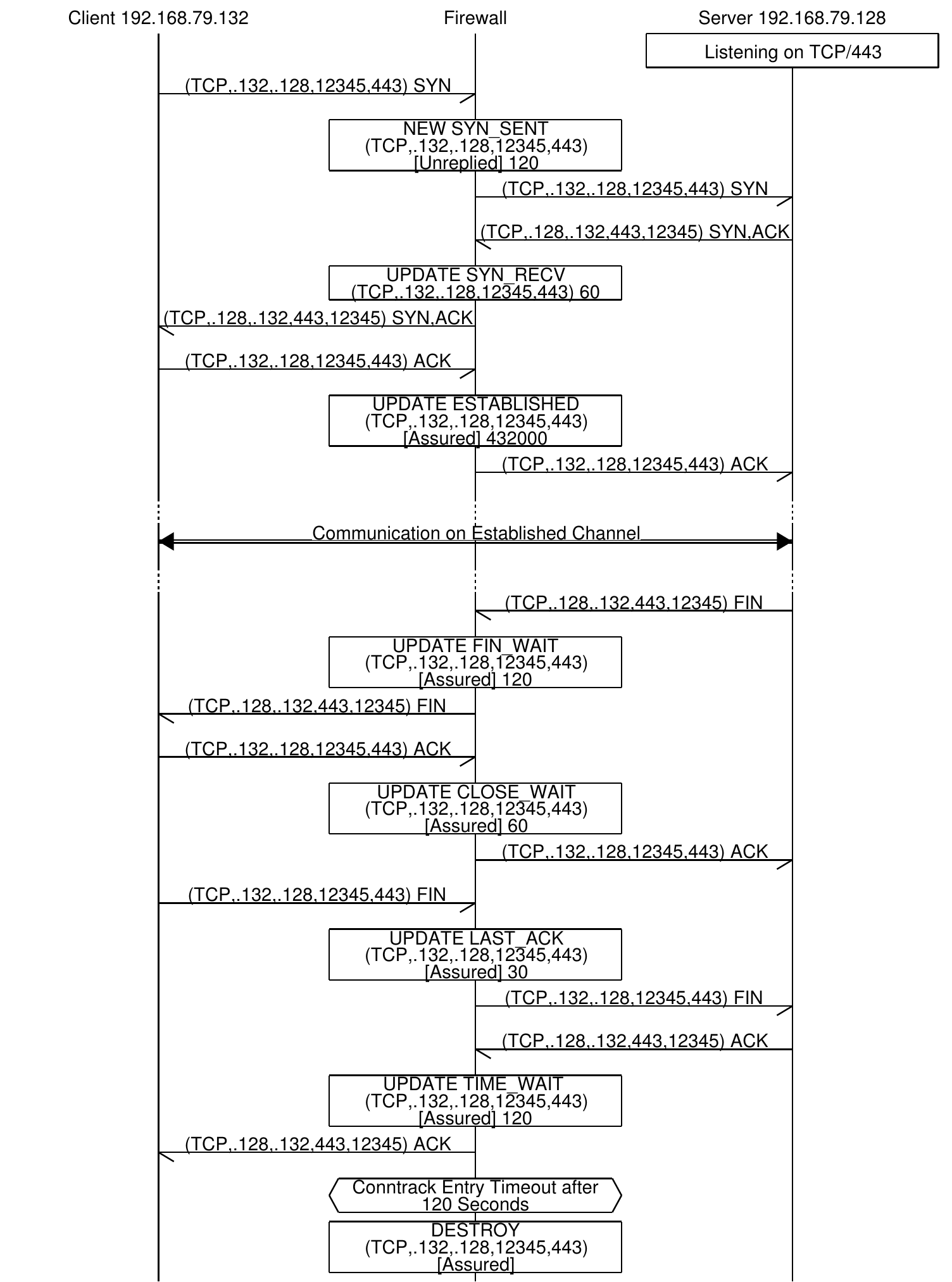}}
	\caption{\emph{Conntrack} events for a TCP packet flow in an HTTP scenario. The 5-tuple is structured as follows: (protocol, src ip, dst ip, src port, dst port).}
	\label{fig:conntrack_table_tcp_handshake}
\end{figure}

\subsection{Stateful TCP}
Figure~\ref{fig:conntrack_table_tcp_handshake} visualizes how TCP packets trigger \emph{conntrack} events and change the connection table. \emph{Conntrack} is able to use the TCP header flags to track the state of a connection. Therefore, new connections from the client always have to start with a \texttt{SYN} packet for which a new table entry is created, storing the 5-tuple of the connection, the state \texttt{SYN\_SENT} and the label \texttt{UNREPLIED}. If the firewall receives a corresponding \texttt{SYN-ACK} from the server the state of the table entry is updated to \texttt{SYN\_RECV} and the \texttt{UNREPLIED} label is removed. After receiving the \texttt{SYN-ACK}, the client sends an \texttt{ACK} message to complete the 3-way handshake. The firewall receives this \texttt{ACK} and updates the table entry to \texttt{ESTABLISHED} and marks the entry as \texttt{Assured}.

Without loss of generality, we assume that the server initiates the teardown of the TCP connection. The teardown is initiated by a \texttt{FIN} message. The firewall updates the connection table entry to \texttt{FIN\_WAIT} and then to \texttt{CLOSE\_WAIT} when the corresponding \texttt{ACK} is received. After receiving the second \texttt{FIN} message from the client, the table entry is updated to \texttt{LAST\_ACK} and finally to \texttt{TIME\_WAIT} as soon as the \texttt{FIN} message is acknowledged by the server. Lastly, the entry times out in the \texttt{TIME\_WAIT} state and is deleted. The purpose for this waiting state is to ensure that packets belonging to the connection are still valid as long as the IP datagrams used in this connection are still valid. In order to reduce the number of packets, modern TCP implementations usually ``piggyback'' the \texttt{ACK} for the first \texttt{FIN} on the second \texttt{FIN} in one combined \texttt{FIN-ACK}. In this case, the firewall skips the \texttt{CLOSE\_WAIT} state and updates the table entry to \texttt{LAST\_ACK} directly.

The other possibility to close a connection is by sending a TCP reset,
which informs the sender to terminate the connection.
Resets should only be used if a received TCP packet cannot be associated with any connection. In order to perform a TCP reset, the \texttt{RST} bit in the header is set. When \emph{conntrack} receives a packet for a matching 5-tuple with a \texttt{RST} bit, the correlating table entry is removed immediately via a \texttt{CLOSE} event.

\subsection{Stateful UDP}
\begin{figure}[tb]
	\centerline{\includegraphics[width=\columnwidth]{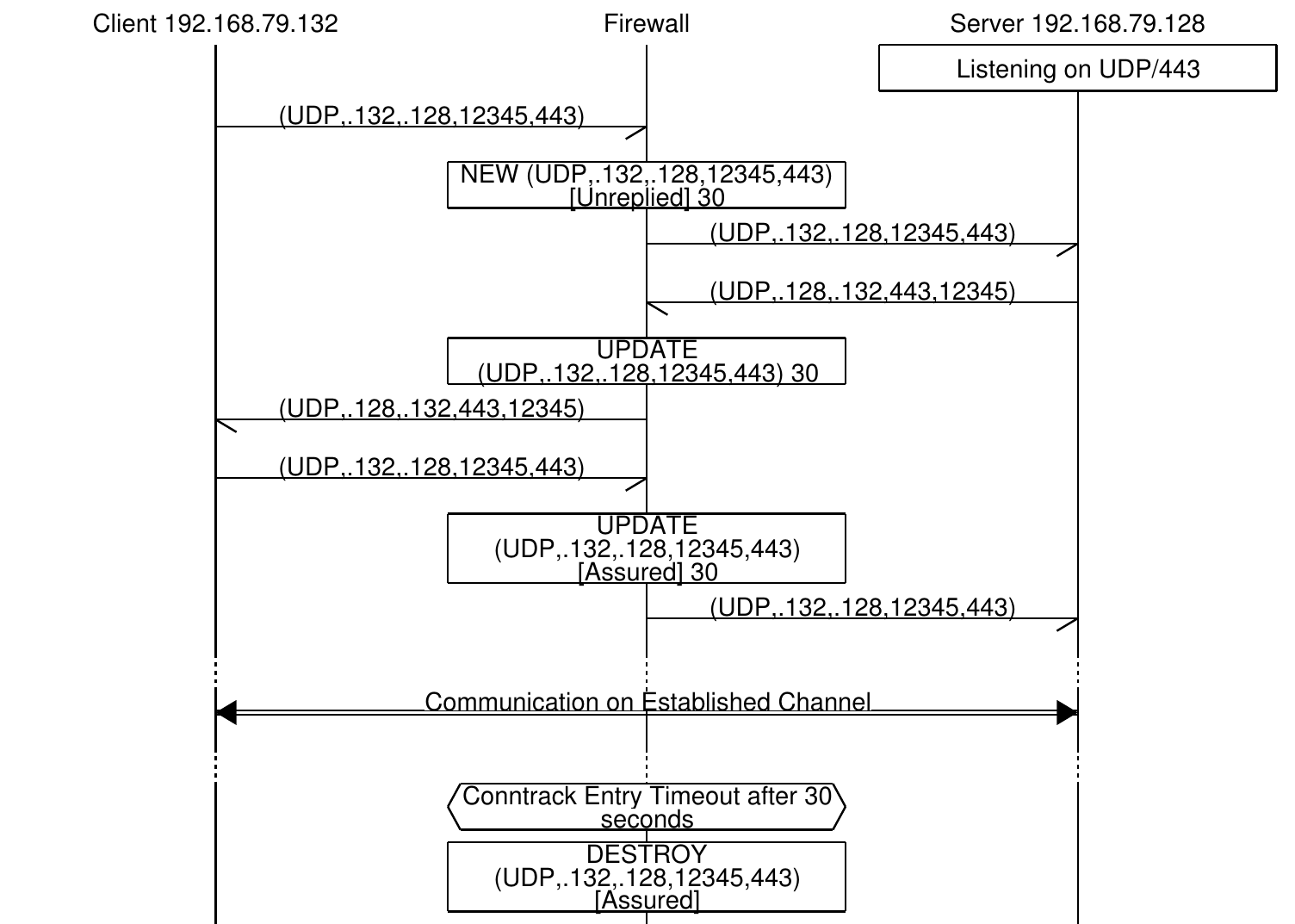}}
	\caption{\emph{Conntrack} events for the UDP packet flow in a HTTP/3 connection.}
	\label{fig:conntrack_table_udp_handshake}
\end{figure}

Figure~\ref{fig:conntrack_table_udp_handshake} shows how \emph{conntrack} attempts to track ``UDP connections''. Since there are no flags representing a connection in UDP packets, the firewall has to treat every UDP packet as a new connection attempt. Thus it creates a new entry for the 5-tuple in the connection table and marks it as \texttt{UNREPLIED}. If the firewall receives a returning packet with the same 5-tuple it assumes that it belongs to the same connection and updates the table entry by removing the \texttt{UNREPLIED} status. For any further packet from the client with the same 5-tuple, the table entry is updated to \texttt{ASSURED}. While it is possible to use large TTL values for established TCP connections because the firewall can use the \texttt{FIN} and \texttt{RST} packets to determine when to remove the connection table entries, entries for UDP have to be updated with each packet to reset the TTL value. That is, UDP is connectionless. There is no teardown or reset procedure and the entries always have to time out.

\subsection{UDP Hole Punching}
Since UDP can only be tracked packet by packet, stateful firewalls are prone to bypass techniques utilizing this behavior. One of the most prominent bypass techniques is UDP hole punching. For our investigation of UDP hole punching in the case of \emph{conntrack} firewalls handling QUIC, we assume that an attacker has already compromised the web server, e.g., via a remote code execution~(RCE) vulnerability on the application level or that she has access to another machine in the same internal network as the web server.

As described above, UDP tries to mimic the TCP 3-way handshake. However, since the packets cannot be associated with a connection, any packet from the server side with the correct 5-tuple subsequently ``punches a hole'' in the firewall for all packets coming from the server side with the same 5-tuple. This technique stems from the core design of UDP and can usually not be avoided. There are even multiple benign applications of this bypass technique in modern networks, e.g., for dealing with network address translation~(NAT) in peer-to-peer~(P2P) networks or for virtual private network~(VPN) connections~\cite{halkes2011}. However, it is also possible to utilize the bypass techniques for malicious purposes.

\begin{figure}[tb]
	\centering
	\includegraphics[width=\columnwidth]{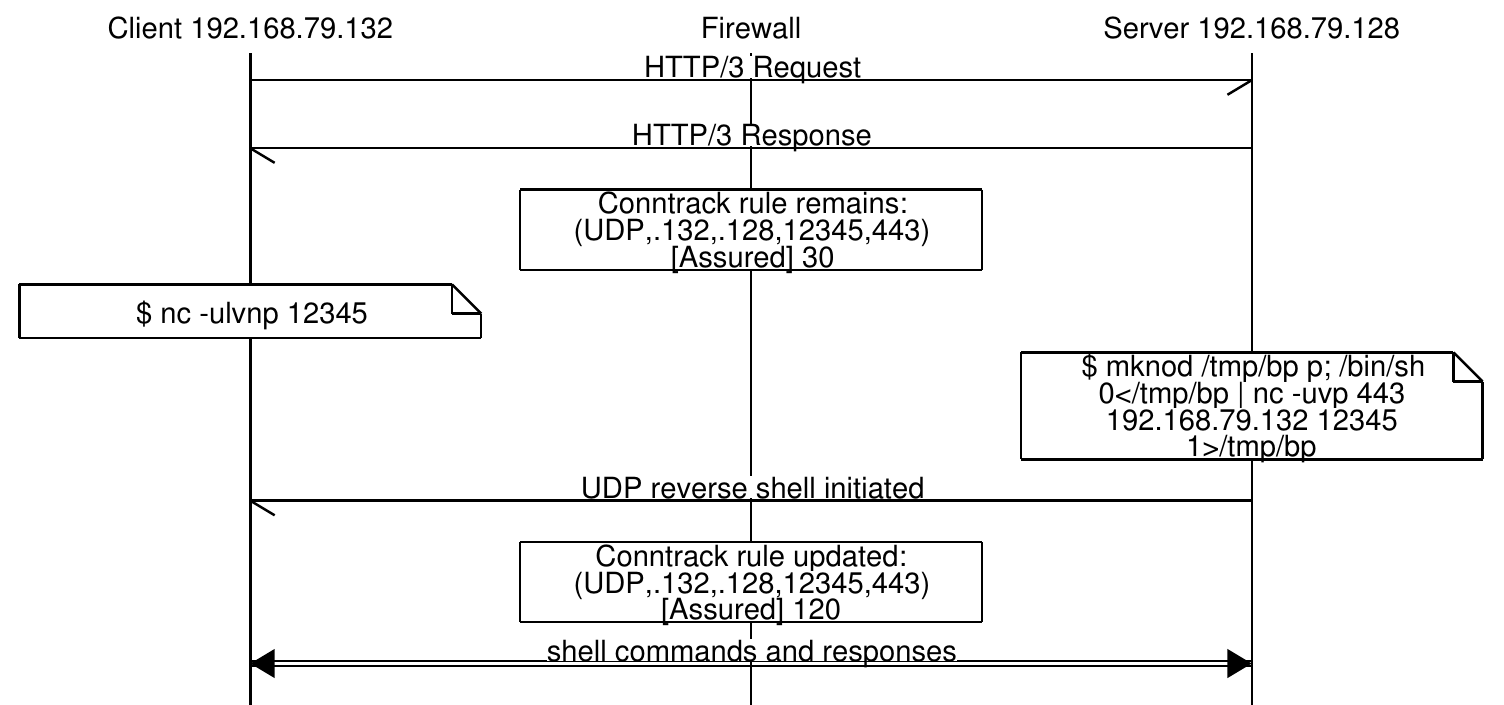}
	\caption{Utilizing UDP hole punching to spawn a reverse shell.}
	\label{fig:udp_reverse_shell}
\end{figure}

\begin{figure*}[!t]
	\centering
	\includegraphics[width=\textwidth]{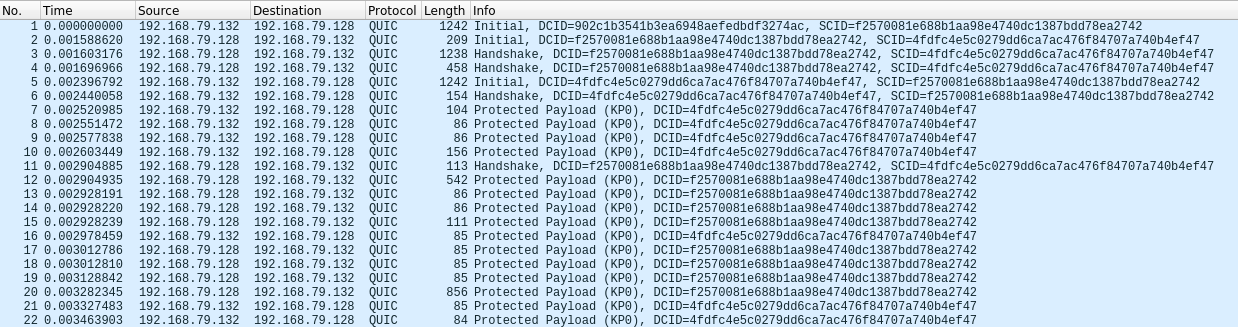}
	\caption{\emph{Wireshark} packet capture for a HTTP/3 request and response.}
	\label{fig:quic_wireshark_capture}
\end{figure*}

The missing teardown for QUIC increases the risk of a UDP hole punching attack being performed successfully. Since the \emph{conntrack} entries for UDP have a TTL of 30 seconds, it is possible for an attacker to indefinitely keep the hole open by sending a UDP packet every few seconds. Therefore, an attacker has an arbitrary amount of time to perform malicious actions even after the original connection has been shut down. While such a hole can always be used to exfiltrate data with a forged IP address and port, an attacker with sufficient control over either the ports on the web server or the IP address mapping to another attacker controlled machine behind the same firewall, e.g., over ARP spoofing, is able to perform more sophisticated attacks. Such an attack is demonstrated in Figure~\ref{fig:udp_reverse_shell} which shows how UDP hole punching though HTTP/3 can be utilized to create a fully interactive reverse shell. The \texttt{TIME\_WAIT} state for TCP also has to time out but the firewall will not accept SYN packets with the same 5-tuple during this period. Therefore, it is not possible to initiate a new connection from the server side during this state that can be exploited in a similar way to the UDP scenario.

\section{QUIC / HTTP/3 Dissection}
\label{sec:quic_http3_dissection}
Since UDP is connectionless, a firewall that aims to provide the same level of connection tracking for QUIC as for TCP, has to analyze the application layer. In general, an application layer firewall would require parsing the whole packet and performing heuristic analysis and string matching operations to determine the protocol, as the application layer protocol is a priori unknown to all middleboxes in a connection. Both the parsing as well as the heuristic analysis would impact the performance in comparison to a transport layer based stateful firewall~\cite{goralski2009}. However, an application layer firewall that only expects QUIC communication can directly parse and interpret the packet, increasing the performance as a result. A real-world performance analysis of different concepts of firewalls for QUIC, it is out of the scope of this paper but should be revisited in the future as soon as implementations are available. In the following sections, we examine which options exist to provide connection based stateful interpretation of QUIC.

\begin{figure}[tb]
	\centering
	\includegraphics[width=\columnwidth]{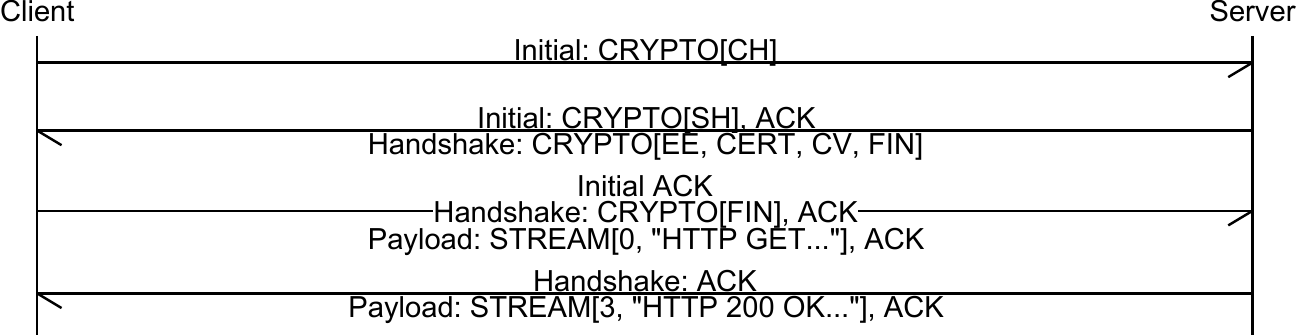}
	\caption{QUIC handshake according to the IETF specification draft with the adapted TLS 1.3 handshake.The first two \texttt{CRYPTO} frames in the initial message carry the \emph{Client Hello} and the \emph{Server Hello} while the \texttt{CRYPTO} frames in the first two handshake messages carry the cipher suite, certificate, signature and finished hashes.}
	\label{fig:quic_hanshake}
\end{figure}

\subsection{Analysis of QUIC traffic with Wireshark}
To analyze QUIC, we have captured the traffic of a complete HTTP/3 request and response in the point-to-point topology described in Section~\ref{sec:setup} with \emph{Wireshark}. \emph{Wireshark} implements heuristic matching steps for application layer protocols as described above and is thereby suitable to represent a hypothetical firewall with deep packet inspection. The payload of a UDP packet is passed to fitting protocol dissectors which parse the packet and perform string matching to check if it could be their protocol. Based on these results \emph{Wireshark} decides to which protocol the packet most likely belongs to. The complete packet capture is shown in Figure~\ref{fig:quic_wireshark_capture}. \emph{Wireshark} is able to distinguish between three types of QUIC messages, \emph{Initial}, \emph{Handshake}, and \emph{Protected Payload}.

Before having a closer look into the individual packet types, the first question that comes up concerns the number of packets captured. QUIC promises to reduce latency by removing RTTs which on a first glance contradicts the 22 packets captured in Figure~\ref{fig:quic_wireshark_capture}. To consider: a comparable HTTPS request over TCP ``only'' needs 18 packets. However, analyzing the source and destination IP addresses reveals that multiple packets are sent in the same direction in bursts which do not affect the number of RTTs. The QUIC specification~\cite{ietfQUIC29} suggests that multiple QUIC frames in the same direction are packed into one UDP packet, if possible while adhering to UDP size limitations. Figure~\ref{fig:quic_hanshake} visualizes the QUIC handshake as proposed in the IETF specification and shows that in total four UDP packets would be sufficient to complete the handshake. However, it is generally allowed to pack every QUIC frame into a separate UDP datagram, as it was implemented in the \emph{quiche} version used. In Figure~\ref{fig:quic_wireshark_capture}, we see that, e.g., the initial frames and the handshake frames from the server are split into the two different UDP datagrams 2 and 3.

\subsection{Unencrypted Content in QUIC Messages}
\label{subsec:unencrypted_QUIC}
As soon as the cryptographic handshake is completed, QUIC only uses a short unencrypted header that contains barely more than a destination connection ID~(DCID). Since this value is intended to track connections for correct delivery when network and transport layer endpoints change, it can also be utilized for connection tracking in a stateful firewall. The first set of connection IDs is always negotiated in the initial and handshake messages visible to an external observer. However, the QUIC specification allows for changing connection IDs via encrypted \texttt{NEW\_CONNECTION\_ID} frames during a connection. As a firewall is not able to decrypt these messages, changes to the DCID would appear spontaneously and could yet only be correlated by matching classical 5-tuple values. It was not possible to reproduce the scenario with the used \emph{quiche} implementation. The DCID set in the initial handshake, stayed unchanged even for continuous connections that lasted longer than 15 minutes. If a final implementation preserves connection IDs in the same way, the DCID value is the most promising option for tracking QUIC connections. Changing DCIDs may still provide some tracking capabilities but they would require a firewall to allow all QUIC packets from one host with a specific 5-tuple regardless of the connection ID. While such behavior should be considered in future work to gain a more in-depth understanding, the bypass capabilities through changing connection IDs appear limited.

The long header of the initial and handshake messages contain, besides an identifier for the packet type, the connection IDs for source and destination (and their lengths). This can be utilized by a firewall, since the DCID of the next returning packet is known in advance. If the IDs are chosen randomly by the peers, it should be infeasible to forge them~\cite{ietfQUIC29}. Furthermore, the structured behavior of the handshake can be used similar to how \emph{conntrack} handles the TCP 3-way handshake to determine the state of a connection.

\subsection{Mapping of the QUIC Handshake to Conntrack States}
All handshake messages are detectable by the long header indicating the initial or handshake packet type. The firewall states for a TCP 3-way handshake can be mapped one-to-one to the inital messages.
\begin{itemize}
	\item A firewall accepts a new connection only, if it is a valid initial packet. It stores the source connection ID~(SCID) and its length to identify following packets to the client together with the 5-tuple of the UDP packet. The state of the connection is marked as \texttt{SYN\_SENT} and \texttt{Unreplied}.
	\item When the firewall receives a returning initial packet, it matches the 5-tuple of the UDP datagram and checks, if the DCID is equal to the SCID of the first initial packet. The SCID of the packet is added to the connection state table together with its length to identify messages to the server. The state is updated to \texttt{SYN\_RECV} and the \texttt{Unreplied} tag is removed. From this point onward, the firewall matches the connection IDs for all packets.
	\item On receiving the next valid and matching initial message from the client, the state is set to \texttt{ESTABLISHED}.
\end{itemize}
The firewall could also keep track of the cryptographic handshake as well as checking its validity. However, this would go beyond tracking only the stream abstraction. In addition to the connection IDs and the extended information in the long header of initial and handshake messages, QUIC exposes no more valuable information for connection tracking. So far, there is still no possibility to track the connection teardown, since all of the involved messages are fully encrypted. QUIC has an equivalent to the TCP reset called stateless reset which is technically not encrypted but requires information only exchanged encrypted and should not be distinguishable from encrypted payload packets. We discuss the stateless reset design and its implications for DoS mitigation in the following section.

\subsection{The Upside to an Encrypted Communication}
\label{subsec:upside_encryption}

On the contrary to the security challenges which the protocol change to UDP inherits, the encrypted connection teardown also has direct benefits in mitigating common strategies for firewall censoring mechanisms. The Great Firewall of China~(GFWC) for example uses, besides other techniques, TCP reset attacks to terminate existing TCP connections. The advantage of terminating connections after a short while over blocking connections in advance is that the communication can be analyzed to gain more insight over the communicating hosts. TCP reset attacks work by forging TCP packets with a RST flag set. If the forger is able to set all values of the RST packet to a convincing value, an endpoint will terminate the connection because they think the packet came from its peer~\cite{weaver2009}.

In theory, there is an equivalent reset attack described in the QUIC specification called ``stateless reset oracle''. However, the feasibility of this attack depends on the attacker being able to generate a 16-byte stateless reset token only known to the endpoints of a connection. These tokens are unique to a connection ID, they should be infeasible to guess and they are issued via encrypted frames in the handshake or during the ongoing communication~\cite{ietfQUIC29}.

A stateless reset packet has the following structure:
\begin{lstlisting}[language=bash,frame=none]
Stateless Reset {
	Fixed Bits (2) = 1,
	Unpredictable Bits (38..),
	Stateless Reset Token (128),
}
\end{lstlisting}
With the first two bits set to \texttt{01} and than at least 38 random bits (in total at least 5 bytes), the packet looks like a typical QUIC packet with a short header and some DCID to an external observer. Since the stateless reset token is only known to the endpoints, it will appear like arbitrary encrypted content. To test the server's reaction to stateless reset packets, we wrote a script using RAW sockets to forge these packets with different possible values as the reset token. However, these packets were not even processed by the server, since the stateless reset feature was not yet implemented in our quiche version. A thorough analysis of this attack should be re-visited for future releases. Even if we were not able to experimentally verify the stateless reset oracle attack, a correct QUIC implementation should leave an attacker no choice but to brute force the reset token which should be infeasible on modern hardware.

\section{Discussion and Future Work}
\label{sec:futurework}
Once more we would like to emphasize that all of the analysis was performed on an experimental implementation of a draft protocol design. Therefore, it is advisable to re-visit the experiments for the final IETF specification, especially the stateless reset oracle attack. It would also be interesting to re-visit some of the work performed for Google's original QUIC design (gQUIC) for QUIC like~\cite{lychev2015, nalawade2018, jager2015}. While the stateful tracking proposed in this work achieves a desirable trade-off between security and censorship resistance, it will likely need more resources than a traditional firewall like \emph{conntrack}. To estimate the real-world feasibility of our approach, it is necessary to conduct a performance comparison and evaluation. Finally, as described in Subsection~\ref{subsec:unencrypted_QUIC}, the QUIC specification allows for DCIDs to change during a connection. While we could not observe this behavior in the used implementation, our stateful tracking approach needs to be re-evaluated for implementations supporting this feature. If implementations are not able to use changing connection IDs, this would immediately expose stateless reset packets because they use a random value as a ``fake'' DCID. Besides the stateless reset oracle, the security considerations in the QUIC specification~\cite{ietfQUIC29} mention other possible attack vectors. According to the number of DoS vulnerabilities mentioned in the specification that could be possible for QUIC, DoS capabilities in general should be analyzed in future research.

\section{Conclusion}
\label{sec:conclusion}
The QUIC protocol inherits many long-needed updates to the way transport protocols are designed today with a strong requirement for security and privacy while reducing latency issues. However, this paper shows that encrypting the stream abstraction of a connection, also limits the capabilities of firewalls to track connection states, introducing new challenges for network operators and firewall developers. On the other hand, the same components limiting these capabilities inherit desirable censorship resistance. Our work provides insight into the limited capabilities within the QUIC protocol design that can be utilized to achieve a stateful firewall design in presence of QUIC traffic. The resistance against reset attacks in contrast to the more difficult state tracking in QUIC shows that the capabilities of firewalls are always a two-sided coin. Achieving more privacy might also impact the possibilities of ``good'' firewalls to keep networks safe.

\IEEEtriggeratref{8}
\bibliographystyle{IEEEtran}
\bibliography{literature}

\end{document}